\documentclass{article}

\usepackage{graphicx}
\usepackage{graphicx}
\usepackage{textcomp}
\usepackage{enumitem}
\usepackage{xcolor}
\usepackage{listings}
\usepackage{booktabs}
\usepackage{longtable}
\usepackage{array}
\usepackage{subcaption}
\usepackage{arydshln}
\usepackage[backend=biber, style=numeric]{biblatex}
\usepackage[utf8]{inputenc} 
\usepackage[T1]{fontenc}    
\usepackage{hyperref}       
\usepackage{url}            
\usepackage{booktabs}       
\usepackage{amsfonts}       
\usepackage{nicefrac}       
\usepackage{microtype}      
\usepackage{xcolor}         
\usepackage{comment}
\usepackage[margin=1in]{geometry}
\usepackage{multirow}
\usepackage{tabularx}
\usepackage{booktabs}
\usepackage{lineno}
\usepackage{titling}
\usepackage{lmodern}
\usepackage{setspace}
\usepackage{titlesec}

\pretitle{%
  \centering
  \rule{\textwidth}{1pt} \\[\bigskipamount]
  \bfseries\LARGE
}
\posttitle{%
  \\[\bigskipamount] \rule{\textwidth}{1pt}
}

\title{A Code Comprehension Benchmark for Large Language Models for Code}

\linenumbers
\nolinenumbers  

\date{}

\author{
  \makebox[\textwidth][c]{%
    \begin{tabular}{ccc}
      \Large\textbf{Jayant Havare\textsuperscript{*}} & \Large\textbf{Saurav Chaudhary\textsuperscript{*}} & \Large\textbf{Ganesh Ramakrishnan} \\
      \large IIT Bombay & \large IIT Bombay & \large IIT Bombay \\
      \small \texttt{} & \small \texttt{} & \small \texttt{} 
    \end{tabular}
  }\\[1em]
  \makebox[\textwidth][c]{%
    \begin{tabular}{cc}
      \Large\textbf{Kaushikraj Maharajan} & \Large\textbf{Srikanth Tamilselvam} \\
      \large IIT Bombay & \large IBM Research \\
      \small \texttt{} & \small \texttt{}
    \end{tabular}
  }
}
\begin{document}

\maketitle
\renewcommand{\thefootnote}{\fnsymbol{footnote}} 
\footnotetext[1]{Both authors contributed equally to this work.}

\begin{abstract}

  Large Language Models have shown impressive capabilities in coding tasks like code generation and code completion, as they have been trained on a large amount of code data. Also, since one of the core pretraining objectives is \textit{Next Token Prediction}, these models tends to learn surface-level syntactic patterns in code. However, this does not guarantee code comprehension ability i.e. the ability to capture the semantics of the code. In our opinion, this is the reason why these models often underperform on tasks that require deeper semantic understanding, such as code debugging and code optimization. To address this, we propose fine-tuning these models specifically for code comprehension tasks using large-scale datasets, enabling them to develop a more robust understanding of code semantics.
  We evaluate three code models of varying sizes on a suite of code comprehension tasks designed to assess semantic understanding beyond surface-level syntactic pattern matching. In particular, we analyze performance on the Subjectivity Grading Task and observe that model performance improves after fine-tuning on relevant downstream tasks. The most significant improvement is seen in the QWQ-32B model, where accuracy increases from 70\% to 83.47\%. A similar or explainable trend is observed across other models, clearly indicating an enhancement in code comprehension ability. Among the models studied, the DPO-fine-tuned Codestral-22B achieves the highest micro-accuracy of 87.66\% on the Subjectivity Grading Task.
\end{abstract}

\section{Introduction} \label{sec:introduction}
Large Language Models (LLMs) are advanced deep learning models trained on vast and diverse datasets, enabling them to generate human-like text. Beyond natural language processing, LLMs have shown impressive capabilities in code generation, translating natural language prompts into syntactically correct programming code. They perform well on standard code generation benchmarks such as HumanEval \cite{chen2021evaluating} and MBPP \cite{austin2021program}. In addition, they are also effective at other code-related tasks, including code completion, code search, code summarization, and code snippet extraction.

\begin{figure*}[t]

\begin{subfigure}{.55\linewidth}

\includegraphics[width=\linewidth, height= 2.6 in]{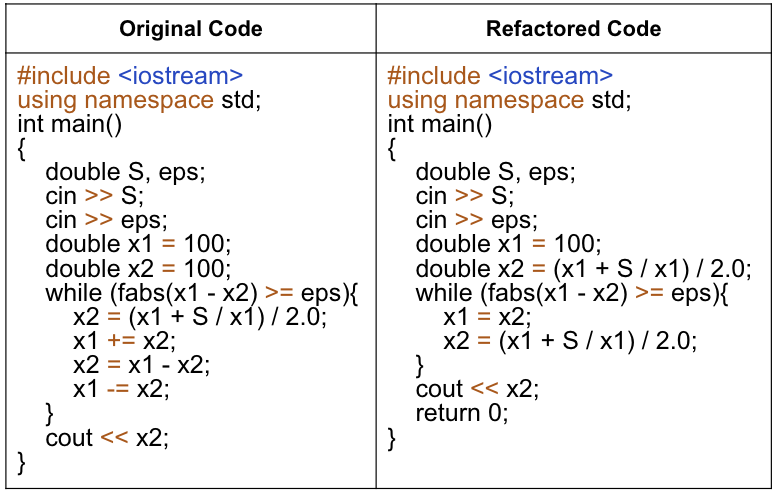}
  \caption{Code Refactoring Task}
  \label{fig:refactor}
\end{subfigure}
\begin{subfigure}{.4\linewidth}
\includegraphics[width=\linewidth, height= 2.6 in]{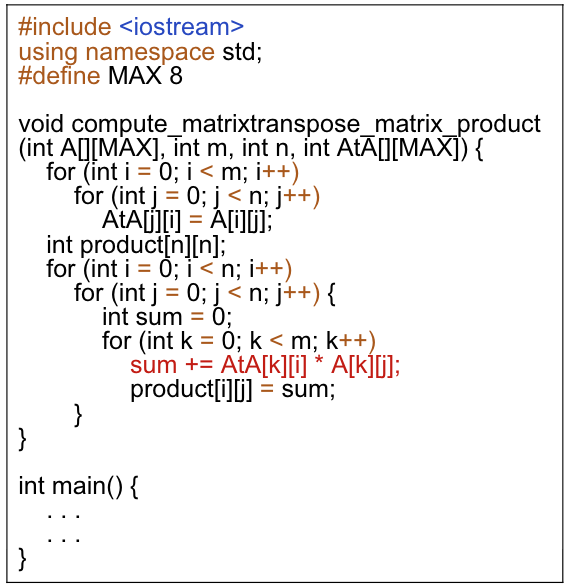}
  \caption{Bug Localization Task }
  \label{fig:bug}
\end{subfigure}

\caption{Semantics-based Tasks}
\label{fig:allgraphs}
\end{figure*}

To evaluate their capabilities in semantics-oriented tasks, we selected \textbf{Codestral-22B}, a high-performing model that has reported strong results on existing benchmarks such as HumanEval ( 81.1\% pass@1-score) and MBPP dataset (78.2\% pass@1-score). 
In Figure~\ref{fig:refactor}, we present both the original and refactored versions of a code that implements the \textit{Babylonian Method}. Here, the model was provided a clear prompt- "Analyze the following code and refactor it to improve readability. Do not change its functionality or output. Ensure that variable names, comments are consistent." However, in the refactored version, the model altered the order of operations within the loop. Specifically, it updated x1 before calculating the new value of x2, thereby breaking the iterative dependency that is critical for convergence. So the model failed in this task. Fig ~\ref{fig:bug} shows another example where the model was prompted with- "\#\#\#System Prompt: Consider the following question and answer the criterion using the provided code. \#\#\#Question: \{“Transpose Matrix times Matrix Product” full length question\}. \#\#\#Criterion: check whether the code correctly implements the transpose matrix times matrix product. \#\#\#Code: \{code\}". In this case, the code included in the prompt contained an error in the computation of the sum variable. Specifically, the line intended to perform the matrix multiplication was incorrect. The correct implementation should have been: sum += AtA[i][k]*A[k][j];. Despite this mistake, the model incorrectly judged the code to be correct, highlighting a failure in its semantic understanding of the task. 

Code comprehension refers to the ability to understand both the functionality and semantics of source code. It involves not only recognizing what the code does, but also understanding its structure and execution flow. This capability is crucial for a variety of downstream tasks such as debugging, refactoring, static analysis, and code optimization.
Despite recent advancements in large-scale code models, most existing models are primarily pre-trained on objectives such as code generation, with "next token prediction" being one of the main tasks. However, this objective does not ensure true code comprehension, which requires an understanding of the structural aspects of code, such as its representation through Abstract Syntax Tree (ASTs) and the flow of the code using Data Flow Graphs (DFGs) or Control Flow Graphs (CFGs). Code comprehension is essential for code models, as it enables LLMs to understand the functionality and structure of existing code, identify potential issues, and perform code-related tasks effectively and efficiently. Incorporating these structural aspects is not only beneficial but essential for developing robust, generalizable, and reliable code models. Since the deep learning models are heavily data-driven, training the model with the code's structural data can improve its comprehension ability.

In this paper, we start the discussion with the prior work done on code comprehension and various approaches explored. Later, we focus on our work by explaining the code comprehension tasks that we selected and that will benefit the model in developing the ability to better process code. Then, we discuss the details of the experiments performed, including the models, their architecture and pretraining tasks, and experimental dimensions. Finally, we discuss the achieved results and compare the performance of the three models across various tasks.

\section{Related Work} \label{sec:related_work}

In the early times, a lot of work was put into the code generation model. Initial approaches involve rule-based and probabilistic systems such as context-free grammars ~\cite{joshi2003formalism}. Also, the Statistical Machine Translation approach was explored by \cite{de2008z3}, which was inspired by Machine Translation Models.

Later on, many deep learning based code generation models were explored. The work of \cite{mastropaolo2021studying} is one of the first to use a transformer-based system for code generation. Code generation progressed with the development of OpenAI's Codex, Google's Alphacode~\cite{li2022competition}, StarCoder~\cite{li2023starcoder}. Those models have shown impressive capabilities in code generation and code completion. \cite{adeli2020supporting} highlight the evolving nature of code comprehension in the context of LLMs, emphasizing the dual importance of traditional reading skills and emerging abilities like prompt engineering. To enhance the coding abilities, the focus was shifted to code comprehension. \cite{gardner2018neural} tried to parse code semantics using LSTMs. \cite{ben2018neural} focused on the use of embeddings over an Intermediate Representation (IR) that leverages contextual flow based on both data- and control-flow. Inspired by Word2Vec, \cite{bertolotti2023fold2vec} captures the semantics using an idea of embeddings. And \cite{fekete2020comprehensive} provides a review of the cognitive software comprehension models. \cite{singh2024empirical} explored code comprehension from a representation learning perspective, leveraging embeddings derived from intermediate representations to capture semantic structure. \\Similarly, benchmarks like HumanEvalPack, LiveCodeBench, and SweLL have been introduced to evaluate code comprehension beyond syntactic correctness, incorporating dynamic program behavior, execution traces, and multi-step reasoning \cite{humanevalpack, livecodebench, swell}.
 These reflect a growing push toward evaluating true semantic understanding over token-level accuracy. These techniques emphasize understanding the control-flow and data-flow of programs, enabling robust comprehension across diverse codebases and programming languages. \cite{middleton2024barriers} examined the process of comprehending code changes during modern code review, identifying key barriers related to understanding context, tooling, and the behavioral implications of refactorings. These studies highlight the need for code comprehension during review tasks. 

Many of the works discussed above address explore the ways to capture the semantics using the structure of the code. To our knowledge, none of them guarantee code comprehension ability because code comprehension requires inherent understanding of the code structure and semantics.

\section{Tasks} \label{sec:tasks}

For evaluation of code comprehension abilities, we need tasks that focus on the structural and semantic aspects of the code. These tasks focus on how well models can reason about functionality, structure, and correctness. We perform a few tasks over multiple dimensions of source code comprehension, including subjective grading, question answering, semantic search, test-case prediction, bug fixing, and code comparison. Each task is supported by a large enough dataset so that models can be trained over multiple aspects of programming knowledge. Next, we describe each task and the corresponding dataset in detail.

\subsection{ Subjectivity Grading Task [SGT]}
The task is as follows: Given (i) a programming problem $P$, (ii) a student’s code submission $S_P$ for problem $P$, and (iii) a rubric criterion $C$ associated with that problem, the model is expected to predict the appropriate rating $R_j \in {R_1, R_2, \ldots, R_{n_C}}$ for $S_P$ with respect to criterion $C$.
To support this task, we introduce \texttt{CS101-Gold}, a curated dataset collected from a previous offering of \textit{CS101}, the introductory programming course at IIT Bombay. The dataset spans a diverse range of programming problems, varying in difficulty and domain. Each problem in the dataset has been \textit{subjectively assessed} using a detailed rubric, which consists of multiple criteria designed to assess different aspects of student understanding and implementation. Each criterion is associated with a discrete set of rating levels that capture the degree to which a submission satisfies the expectations of that criterion. These ratings were assigned manually by teaching assistants and instructors, following a well-defined evaluation rubric. This kind of fine-grained, criterion-based subjective assessment has long been used in educational settings educational setting since a long time because it enables a detailed feedback, better learning and smooth grading. Each data point in \texttt{CS101-Gold} is represented as a tuple: $<P, S_P, C, \mathbb{R}_C>$
where $P$ is the problem statement, $S_P$ is the student's code, $C$ is a rubric criterion, and $\mathbb{R}_C$ is the set of ratings assigned for criterion $C$. The dataset contains a total of 27 Programming Questions, 210 criteria, 3725 submissions, and 27,699 datapoints.



\subsection{Code Question-Answering Task [QAT]}
The task is as follows: Given a source code and a natural language question about code, a free-form textual answer is required to be generated. The textual answer may be a word, a phrase, or a sentence, and it usually cannot be directly extracted from the source code. This task is different from traditional machine reading comprehension, as code is very different from text, and we need programming knowledge to understand it. The source code and the natural language question are actually in two different languages and the model should have the ability to understand both the code language and the natural language. Moreover, the model needs to generate an answer faithful to the question and the corresponding code, rather than extract some tokens from the code. The dataset we use is the CodeQA dataset. it is a free-form question answering dataset for the purpose of source code comprehension: given a code snippet and a question, a textual answer is required to be generated. It contains a Java dataset with 119,778 question-answer pairs and a Python dataset with 70,085 question-answer pairs. 



\subsection{Code Search Task [CST]}
The task is as follows: Given a natural language query, the goal is to extract the most relevant source code-fragment that matches the query. The model requires a deep understanding of both natural language and programming language for this task. The model must learn a mapping between a code and code-query in while training process. This task is crucial for improving developer productivity, enabling intelligent code retrieval from large codebases. The dataset we use is CodeSearchNet. It is a large-scale benchmark dataset designed specifically for semantic code search and related tasks. The dataset provides millions of code snippets along with corresponding natural language documentation in six programming languages. It is widely used for training and evaluating models on code retrieval tasks. The dataset contains 4,12,178 train and 22,176 test Python datapoints.


\subsection{Test Case Task [TCT]}
This is the upgraded version of the dataset used in the Subjectivity Grading Task. Here we have added 5 test-cases for each question, and the task is to predict whether the provided code passes the test-case or not. The dataset used is same as the Subjectivity Grading Task.

\subsection{Bug Fix Task [BFT]}
The task is as follows: Given a buggy source code and a natural language prompt instructing to correct the code, the objective is to identify the error(s) in the provided code and generate a syntactically and semantically correct version that addresses the issue.
We are using the dataset introduced by Tufano et al. \cite{tufano2019empirical}, which consists of pairs of buggy and fixed Java functions. Each data point includes a buggy function as the source and its corresponding corrected version as the target. We first evaluate various large language models (LLMs) in their raw, zero-shot form on the test set. Subsequently, we fine-tune each LLM using the training and validation sets, and re-evaluate their performance on the test set.

To construct a realistic and diverse code-repair dataset, we first transform our rubric-based subjective grading dataset into a paired buggy–fixed dataset. This involves selecting a subset of programming lab assignments from the CS101 course, configuring them on our in-house LMS-X platform with problem statements, test cases, and student submissions. These submissions are evaluated using an autograder, and based on the results, are categorized as either functionally correct (fixed code) or incorrect (buggy code), resulting in a dataset of 5,336 submissions. To further augment the dataset and improve model generalization, we generate additional synthetic buggy–fixed pairs using semantic similarity techniques. For each correct (fixed) submission, we compute its code embedding using GraphCodeBERT and retrieve the top five most semantically similar buggy submissions from the dataset. This similarity-based pairing process yields an additional 20,096 buggy–fixed examples.

\subsection{Code Comparison Task [CCT]}

The task is as follows: Given two student source codes written as solutions to a particular programming problem, we aim to distinguish the student code which has better alignment with the highest rating of a particular criterion of the programming problem. The target answer is either “code\_1” or “code\_2”. This task tests the model’s ability to map a code to a particular natural language (criterion, rating pair) effectively. For training the model, we construct the training data indigenously from our own graded dataset and do supervised finetuning on a pre-trained LLM. The dataset used is derived from the Subjectivity Grading Task's dataset and contains 748,940 datapoints.

\section{Experiments} \label{sec:experiments}
LLMs with billions of parameters demand extensive computational resources that lead to significant energy consumption during both training and testing. This high energy usage raises concerns about sustainability, as it contributes to increased carbon emissions and environmental impact. Therefore, focusing on smaller models while maintaining the performance is essential. This motivates us to minimize GPU usage to reduce both computational cost and environmental impact. That's why we focus our evaluations on models up to approximately 30 billion parameters, maintaining a balance between performance and resource efficiency. To demonstrate the generalizability of our approach, we evaluate three models ranging from 8 billion to 32 billion parameters. This range allows us to prove that code comprehension capabilities hold consistently across different model sizes. We perform semantics-based tasks discussed in the section \ref{sec:tasks}, considering the different models and experimental dimensions as explained next.

\subsection{Models}

We have chosen three different models, considering a wide range with respect to the number of parameters, which are discussed below:

\begin{enumerate}
    \item \textbf{QWQ-32B [Q]}\\
    \textbf{QWQ} stands for Qwen with Questions. It is Alibaba’s AI reasoning model that prioritizes analytic reasoning and computational efficiency. Built upon the Qwen 2.5-32B architecture, it is a decoder-only transformer with 32 billion parameters and supports a context length of 32K tokens. Diverging from traditional extensive pretraining strategies, its training paradigm emphasizes enhancing reasoning capabilities during inference. It demonstrates strong performance in mathematical and programming reasoning tasks, matching the capabilities of models like DeepSeek R1 while using fewer computational resources. It leverages reinforcement learning (RL) to specialize further—using outcome-based rewards such as accuracy verifiers for math and test case-based verifiers for code. Additionally, RL is employed to improve general language capabilities, resulting in better instruction following and improved alignment with human preferences. The model’s effectiveness has been validated on various benchmarks, including AIME24, LiveCodeBench, IFEval, and BFCL.
    
    \item \textbf{Codestral-22B [CS]}\\
    Mistral AI entered the domain of LLMs for code with its first model called \textbf{Codestral}, showcasing impressive results across multiple benchmarks. It is an open-weight generative AI model. “Open-weight” means that the model parameters are freely available for research purposes. Codestral-22B is a decoder-only transformer with 22 billion parameters, trained on a diverse dataset encompassing over 80 programming languages, and demonstrates multilingual capability in coding tasks. It employs both next-token prediction and a fill-in-the-middle (FIM) mechanism to generate and complete code snippets with high accuracy. The model has outperformed larger counterparts such as Meta’s Llama3 70B and DeepSeek Coder 33B on HumanEval FIM benchmarks. Its capabilities have been validated across a variety of datasets, including HumanEval, MBPP, CruxEval-O, and RepoBench, establishing Codestral-22B as a strong performer in code-centric AI applications.

    \item \textbf{Granite-8B [GR]}\\
    This model is part of IBM’s \textbf{Granite} series, a decoder-only transformer model designed to enhance code intelligence and support enterprise software development. With 8 billion parameters and an extended context window of up to 128K tokens, it is particularly well-suited for long-context understanding in real-world development environments. The model was trained in two phases: the first phase involved 4 trillion tokens from 116 programming languages, building a robust understanding of syntax and code semantics; the second phase included 500 billion tokens from a mix of code and natural language, improving its reasoning and instruction-following abilities. Optimized for enterprise workflows, Granite-8B excels in tasks such as code generation, bug fixing, code explanation, and documentation. It delivers state-of-the-art performance among open-source code models and integrates seamlessly into professional software development pipelines. Its capabilities have been validated on datasets like CodeNet, further highlighting its effectiveness in diverse programming scenarios.
\end{enumerate}

\subsection{Experimental Dimensions}
To holistically evaluate the code comprehension capabilities of LLMs based on the tasks and dataset we are publishing in this work, we define three somewhat complementary evaluation axes: (i) Pre-hoc vs. Post-hoc reasoning, (ii) Intrinsic vs. Extrinsic grounding, and (iii) Abstractive vs. Extractive capability.
These three dimensions capture different perspectives of code comprehension and allow us to explore where each model excels or struggles. This provides a comprehensive view of their code comprehension capabilities. 


\subsubsection{Post-hoc vs. Pre-hoc Evaluation}
One of the key distinctions in evaluating code comprehension lies in whether the task involves analysis of existing code (\textit{post-hoc}) or generation of code before it exists (\textit{pre-hoc}). This distinction helps capture different aspects of a model's capabilities.
\paragraph{1. Post-hoc Evaluation}  
This setting applies to scenarios where the code is already written, and the model must analyze or process it. Such tasks evaluate how well the model can detect errors, understand the functionality of the code, and assess its quality. They reflect real-world applications like code reviews, debugging, or grading student assignments, focusing more on the model’s analytical capabilities than on generation. In our case, we consider the \textit{Subjectivity Grading Task}, where the model analyzes a student’s code submission and selects an appropriate rating for a given rubric criterion.
\paragraph{2. Pre-hoc Evaluation}  
This setting pertains to tasks where the model is required to generate or complete code, often starting from minimal or incomplete input. These tasks assess the model’s ability to understand a problem and produce a meaningful and correct code solution. Examples include code completion or bug-fixing tasks. In our context, we consider a \textit{bug-fix task}, where the model is provided with an erroneous piece of code and must produce a corrected version. This requires the model to infer the intended functionality, identify errors, and generate syntactically and semantically correct code. Such tasks evaluate both the model’s code generation skills and its understanding of programming logic and common error patterns.

\subsubsection{Intrinsic vs. Extrinsic Evaluation}

Another important distinction in evaluating code comprehension is whether the task measures the model’s internal understanding of code (\textit{intrinsic}) or its ability to apply that understanding to achieve a broader goal (\textit{extrinsic}). This categorization helps assess both the structural analysis capabilities of the model and its functional application skills.  

\paragraph{1. Intrinsic Evaluation}  
Intrinsic evaluation focuses on tasks that directly assess how well the model understands the structure, logic, and semantics of code. These tasks are independent of any external objective and evaluate the model’s internal representation of code. For example, we consider the \textit{Code Comparison Task}, where the model is given two code snippets and must determine which one is better. This requires the model to analyze syntax, control flow, and logical correctness.  

\paragraph{2. Extrinsic Evaluation}  
Extrinsic evaluation assesses the model’s understanding of code indirectly—through its performance on goal-oriented tasks. These tasks reflect real-world scenarios where the model must apply its learned code representations. For instance, in a \textit{Test Case Prediction Task}, the model is asked to predict whether a given code will pass a specified set of test cases. Although the goal is outcome-based, successful prediction demands a deep understanding of the code’s behavior and logic, simulating compiler-like reasoning.  

\subsubsection{Abstractive vs. Extractive Evaluation}

A third perspective in evaluating code comprehension involves distinguishing between the model’s ability to generate new insights from code (\textit{abstractive}) and its ability to extract specific information from code (\textit{extractive}). This helps assess both generative reasoning and selective attention capabilities of the model.  

\paragraph{1. Abstractive Evaluation}  
Abstractive tasks require the model to go beyond locating or copying code elements; it must infer meaning and generate novel outputs. These tasks evaluate the model’s deeper understanding of semantics and its ability to explain or summarize code. We consider a \textit{Code Question Answering Task}, where the model answers natural language questions about a given code snippet. Accurate answers depend on the model’s ability to comprehend and abstract the logic into coherent, explanatory responses.  

\paragraph{2. Extractive Evaluation}  
Extractive tasks require the model to identify and retrieve specific information from code, testing its ability to focus on relevant segments while ignoring noise. For example, in a \textit{Code Search Task}, the model is provided with a natural language query (e.g., “extract the code fragment that sorts a list”) and must return the corresponding code segment. This reflects practical use cases such as code navigation and snippet retrieval in development environments.

\section{Results \& Discussion} \label{sec:results}
Now, we discuss the results achieved from all the experiments. We anticipated that these downstream tasks would help models develop the ability to better analyze the code, and the actual results show a similar trend, which is discussed next. As shown in Table \ref{tab:results} (A:micro-accuracy, F:micro-f1 score\cite{micro_f1}, R:ROUGE-score\cite{rouge}, C:CodeBLEU-score\cite{ren2020codebleu}), we have experimented with the different code comprehension tasks over 3 different range models. A general trend of improving the model's comprehension ability can be noticed. Next, we discuss quantitative as well as qualitative analysis. 

\begin{table}[htbp]
\centering
\begin{tabular}{|p{0.5cm}|p{1cm}|p{1cm}|p{1cm}|p{0.8cm}|p{1cm}|p{1.5cm}|p{1.5cm}|p{0.8cm}|p{1cm}|p{1cm}|}
\hline
\textbf{} & \textbf{SGT} & \textbf{CST} & \textbf{CST + SGT} & \textbf{QAT} & \textbf{QAT + SGT} & \textbf{TCT} & \textbf{TCT + SGT} & \textbf{BFT} & \textbf{BFT + SGT} & \textbf{CCT + SGT} \\
\hline
\multirow{3}{*}{\textbf{Q}} 
& A = 70\% & R = 0.49 & A = 75.56\% & R = 0.42 & A = 82.53\% & accuracy = 82.97\% & A = 82.36\% & C = 0.59 & A = 82.49\% & A = 83.47\%\\
& F = 65\% & & F = 71.88\% & & F = 78.14\% & & F = 77.96\% & & F = 78.09\% & F = 81.31\% \\
\hline
\multirow{3}{*}{\textbf{CS}} 
& A = 86\% & R = 0.40 & A = 85.69\% & R = 0.070 & A = 82.04\% & accuracy = 82.4\% & A = 86.14\% & C = 0.41 & A = 87.66\% & A = 83.73\%\\
& F = 82\% & & F = 81.09\% & & F = 77.47\% & & F = 82.42\% & & F = 84.35\% & F = 79.82\%\\
\hline
\multirow{3}{*}{\textbf{GR}} 
& A = 68\% & R = 0.47 & A = 76.88\% & R = 0.077 & A = 76.12\% & accuracy = 41.78\% & A = 75.69\%
 & C = 0.62 &  & A = 76.55\%\\
& F = 58\% & & F = 70.69\% & & F= 70.04\%  
&  & F = 70.01\% &  & NA & F = 69.43\%\\
\hline
\end{tabular}
\caption{Performance comparison across multiple evaluation tasks across three models}
\label{tab:results}
\end{table}
\subsection{Quantitative Analysis}
Here, we analyze the impact of the different tasks on each model one by one, and we try to prove how these tasks help the model to get better at analyzing the code. We also discuss the reasons behind the models' behavior after finetuning.

Initially, QWQ-32B model achieved a 70\% micro-accuracy and a 65\% micro-f1 score on the standalone subjectivity grading task. However, when combined with other tasks, significant improvements were observed. In the Code-Search task, the baseline ROUGE score was 0.49. After finetuning with the Subjectivity Grading Task, micro-accuracy (evaluated over task-specific correctness) improved to 75.56\%, and the micro-f1 score rose to 71.88\%. 
This indicates enhanced retrieval and evaluation capabilities. Similarly, in the Question Answering, starting from a ROUGE score of 0.42, the integration of grading elevated micro-accuracy to 82.53\% and micro-f1 score to 78.14\%, reflecting improved comprehension and response quality. During the test Case Prediction task, the standalone accuracy is 82.97\%, and when combined with grading, micro-accuracy increased to 82.36\%, and micro-f1 score to 77.96\%, suggesting a positive impact on grading task. Finally, in the Bug Fixing task, the initial CodeBLEU score was 0.59, and incorporating grading led to an micro-accuracy of 82.49\% and an micro-f1 score of 78.09\%, demonstrating improved code grading capabilities. Finally, after training on the Code Comparison Task, the model achiveved 83.47\% of micro-accuracy and 81.31\% of micro-f1 score. In addition, we also evaluated this model on Bug Fix on a proposed dataset (Sec. 3\ref{sec:tasks}) which we curated from the Subjectivity Grading dataset, where the evaluation strategy was to check whether model is able to completely fix the code or not (i.e. we consider it correct if the code is completely fixed). Here, we noticed a significant jump of 18\% where initially the model gave an accuracy of 20\% and after finetuning it reached 38\% which again demonstrates enhancement in code comprehension ability.

The Granite-8B model demonstrates significant performance enhancements across various programming-related tasks when subjectivity grading is integrated. Initially, on the standalone subjectivity grading task, it achieved 68\% micro-accuracy and a 58\% micro-f1 score. However, incorporating grading into downstream tasks led to notable improvements. In the Code Search task, micro-accuracy increased to 76.88\% and micro-f1 score to 70.69\%, indicating more precise retrieval. The Question Answering task saw micro-accuracy rise to 76.12\% and micro-f1 score to 70.69\%, reflecting better contextual understanding. For Test Case Prediction, micro-accuracy reached 76.1\% with an micro-f1 score of 70.01\%, suggesting more consistent predictions. In the Bug Fix task, the model's responses were out of asked format in many places, so we could not calculate the results properly. This model has shown the best micro-accuracy on the Subjectivity Grading Task of 76.55\% after being trained on the Code Comparison Task.

While the Codestral model initially achieved much better results—86\% micro-accuracy and 82\% micro-f1 score on the standalone subjectivity grading task—integrating subjectivity grading into downstream tasks resulted in almost the same performance. In the Code Search task, for instance, the baseline ROUGE score was 0.40, and despite the integration of grading, micro-accuracy and micro-f1 scores remained at 85.69\% and 81.09\%, respectively, indicating no significant improvement in retrieval precision. Similarly, in the Question Answering task, starting from a ROUGE score of 0.07, the addition of grading did not improve micro-accuracy and micro-f1 scores beyond 82.04\% and 77.47\%, respectively, resulting limited impact on comprehension. In the test Case Prediction task, the standalone accuracy was 82.4\%, and with grading, micro-accuracy and micro-f1 scores achieved is 86.14\% and 82.42\%, respectively, indicating similar performance as compared to fine-tuned over subjectivity grading task. Finally, in the Bug Fix task, the initial CodeBLEU score was 0.41, and incorporating grading led to an micro-accuracy of 87.66\% and an micro-f1 score of 84.35\%, which are the best achieved results for Subjectivity Grading Task. The micro-accuracy has reduced to 83.73\% with micro-f1 score of 79.82\% after training on the Code Comparison Task. In the codestral's case, the performance has reduced in subjectivity grading after training over some downstream tasks. Several factors may contribute to this lack of improvement. Firstly, there is a chance of catastrophic forgetting where the model is forgetting its pretrained knowledge because it undergoes multiple finetuning tasks. Secondly, the Codestral model's architecture and training might not be optimized to leverage the benefits of finetuning multiple tasks. 

\subsection{Qualitative Analysis}
\begin{figure*}[t]
    \centering
    \begin{subfigure}{.49\linewidth}
        \includegraphics[width=\linewidth, height=2.6in]{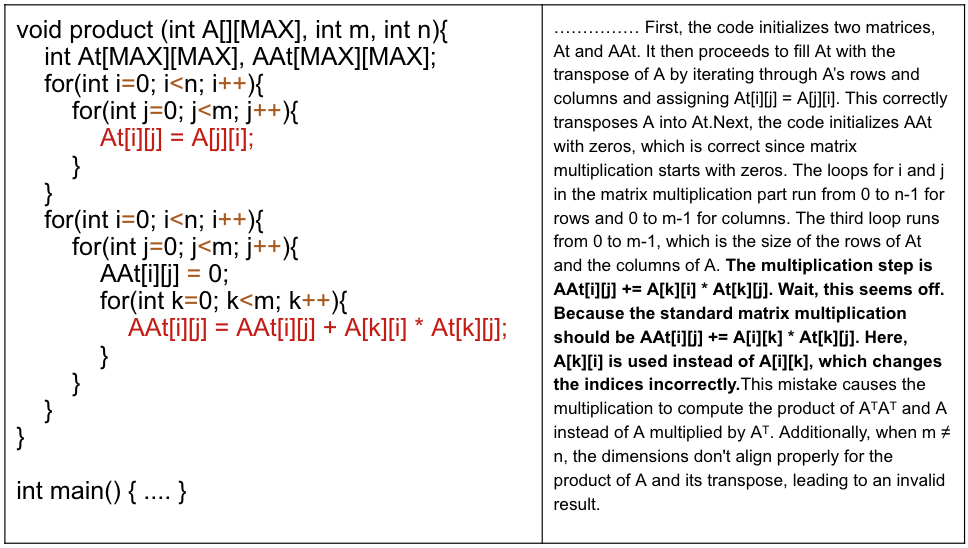}
        \caption{Qualitative Example 1}
        \label{fig:qe1}
    \end{subfigure}
    \hfill
    \begin{subfigure}{.49\linewidth}
        \includegraphics[width=\linewidth, height=2.6in]{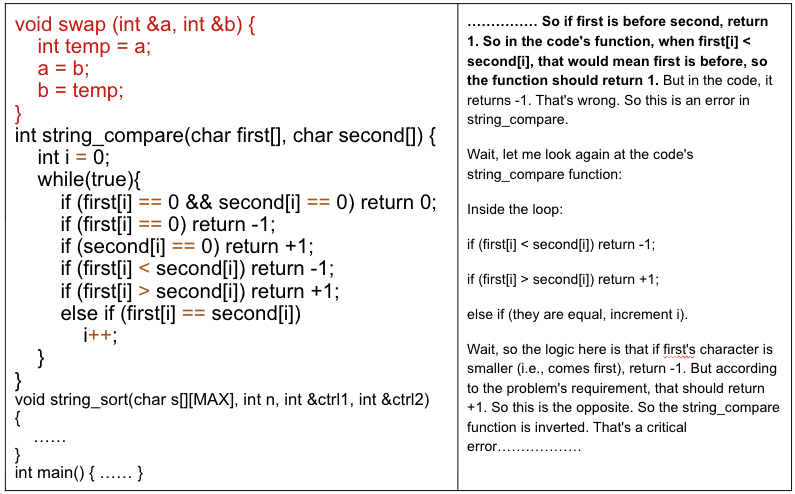}
        \caption{Qualitative Example 2}
        \label{fig:qe2}
    \end{subfigure}
    \caption{Qualitative examples (only the necessary parts of the code and response are shown)}
    \label{fig:qualitative_examples}
\end{figure*}

We further validate code comprehension using qualitative analysis. To our knowledge, the models fail to analyze long or complex codes. So we experimented with subjectivity grading for long and complex codes. 
Figures~\ref{fig:qe1} and~\ref{fig:qe2} illustrate two representative examples from our qualitative analysis. In the first example, the model identifies a non-trivial error in matrix multiplication logic. In the second, the model detects a subtle bug involving string comparison logic—demonstrating improved semantic reasoning.
The first example, shown in Fig. \ref{fig:qe1}, has the code for the programming task``Product of Matrix Transpose times Matrix” on the left, and the model's response on the right. We asked the model to check if the code correctly performs this task. In its response, the model accurately identifies a complex error in the code. Similarly, Fig. \ref{fig:qe2} shows the code for``Insertion Sort over Strings” and the model's response. The model was asked whether the code correctly implements insertion sort for strings. As seen on the right, the model correctly points out a small mistake in how the strings are compared. It also notes that the swap function is written for integers, not strings (as mentioned earlier in its response).

\section{Conclusion} \label{sec:conclusion}
In this paper, we introduced a comprehensive benchmark suite for evaluating code comprehension in large language models. Our results demonstrate that training on semantically rich tasks like grading and bug fixing significantly enhances model comprehension. These findings call for a paradigm shift in the evaluation of code LLMs—from syntactic correctness to semantic understanding—and lay the groundwork for more robust, interpretable code models.
For instance, for the Subjectivity Grading Task, the model QWQ-32B achieved a jump in micro-accuracy from 70\% to 83.47\% which is noticeable and suggests that the model is getting better at code comprehension, and the best performance achieved is 87.66 \% using DPO finetuned Codestral-22B model that was trained over the Bug Fix Task. We are continuing this work, exploring other tasks where models can better capture the semantics, considering the inherent structure of the code. Also, we are exploring ways where we can provide debugger output to models so that they can comprehend the code better.

\printbibliography

@inproceedings{joshi2003formalism,
  title={A formalism for dependency grammar based on tree adjoining grammar},
  author={Joshi, Aravind and Rambow, Owen},
  booktitle={Proceedings of the Conference on Meaning-text Theory},
  pages={207--216},
  year={2003},
  organization={MTT Paris, France}
}

@inproceedings{tufano2019empirical,
  title={An empirical study on learning bug-fixing patches in the wild via neural machine translation},
  author={Tufano, Michele and Watson, Cody and Bavota, Gabriele and Penta, Massimiliano Di and White, Martin and Poshyvanyk, Denys},
  booktitle={ACM Transactions on Software Engineering and Methodology (TOSEM)},
  volume={28},
  number={4},
  pages={1--29},
  year={2019},
  publisher={ACM New York, NY, USA}
}

@inproceedings{de2008z3,
  title={Z3: An efficient SMT solver},
  author={De Moura, Leonardo and Bj{\o}rner, Nikolaj},
  booktitle={International conference on Tools and Algorithms for the Construction and Analysis of Systems},
  pages={337--340},
  year={2008},
  organization={Springer}
}

@inproceedings{gardner2018neural,
  title={Neural semantic parsing},
  author={Gardner, Matt and Dasigi, Pradeep and Iyer, Srinivasan and Suhr, Alane and Zettlemoyer, Luke},
  booktitle={Proceedings of the 56th annual meeting of the association for computational linguistics: tutorial abstracts},
  pages={17--18},
  year={2018}
}

@inproceedings{mastropaolo2021studying,
  title={Studying the usage of text-to-text transfer transformer to support code-related tasks},
  author={Mastropaolo, Antonio and Scalabrino, Simone and Cooper, Nathan and Palacio, David Nader and Poshyvanyk, Denys and Oliveto, Rocco and Bavota, Gabriele},
  booktitle={2021 IEEE/ACM 43rd International Conference on Software Engineering (ICSE)},
  pages={336--347},
  year={2021},
  organization={IEEE}
}

@article{li2022competition,
  title={Competition-level code generation with alphacode},
  author={Li, Yujia and Choi, David and Chung, Junyoung and Kushman, Nate and Schrittwieser, Julian and Leblond, R{\'e}mi and Eccles, Tom and Keeling, James and Gimeno, Felix and Dal Lago, Agustin and others},
  journal={Science},
  volume={378},
  number={6624},
  pages={1092--1097},
  year={2022},
  publisher={American Association for the Advancement of Science}
}

@article{li2023starcoder,
  title={Starcoder: may the source be with you!},
  author={Li, Raymond and Allal, Loubna Ben and Zi, Yangtian and Muennighoff, Niklas and Kocetkov, Denis and Mou, Chenghao and Marone, Marc and Akiki, Christopher and Li, Jia and Chim, Jenny and others},
  journal={arXiv preprint arXiv:2305.06161},
  year={2023}
}

@article{ben2018neural,
  title={Neural code comprehension: A learnable representation of code semantics},
  author={Ben-Nun, Tal and Jakobovits, Alice Shoshana and Hoefler, Torsten},
  journal={Advances in neural information processing systems},
  volume={31},
  year={2018}
}

@misc{humanevalpack,
  author       = {Tong Ye and others},
  title        = {HumanEvalPack: A Multilingual and Diagnostic Expansion of HumanEval},
  year         = {2024},
  howpublished = {\url{https://github.com/tongye98/Awesome-Code-Benchmark}},
  note         = {Accessed: 2025-07-13}
}

@article{livecodebench,
  title        = {LiveCodeBench: A Contamination-Free Evaluation for LLMs on Competitive Programming},
  author       = {Yucheng Lu and Ruiqi Zhong and Danqi Chen},
  journal      = {arXiv preprint arXiv:2403.07974},
  year         = {2024},
  url          = {https://arxiv.org/abs/2403.07974}
}

@misc{swell,
  author       = {RunLoop AI},
  title        = {SweLL Benchmark: Semantic Code Evaluation Beyond Generation},
  year         = {2024},
  howpublished = {\url{https://www.runloop.ai/blog/understanding-llm-code-benchmarks-from-humaneval-to-swe-bench}},
  note         = {Accessed: 2025-07-13}
}

@article{chen2021evaluating,
  title={Evaluating Large Language Models Trained on Code},
  author={Chen, Mark and Tworek, Jerry and Jun, Heewoo and Yuan, Qiming and Pinto, Henrique Ponde de Oliveira and Kaplan, Jared and Edwards, Harri and Burda, Yuri and Joseph, Nicholas and Brockman, Greg and others},
  journal={arXiv preprint arXiv:2107.03374},
  year={2021}
}

@article{austin2021program,
  title={Program Synthesis with Large Language Models},
  author={Austin, Jacob and Odena, Augustus and Nye, Maxwell and Bosma, Maarten and Michalewski, Henryk and Dohan, David and Jiang, Ellen and Chen, Carrie and Zhang, Aitor Lewkowycz and Saunders, William and others},
  journal={arXiv preprint arXiv:2108.07732},
  year={2021}
}

@article{bertolotti2023fold2vec,
  title={Fold2vec: Towards a statement-based representation of code for code comprehension},
  author={Bertolotti, Francesco and Cazzola, Walter},
  journal={ACM Transactions on Software Engineering and Methodology},
  volume={32},
  number={1},
  pages={1--31},
  year={2023},
  publisher={ACM New York, NY}
}

@inproceedings{fekete2020comprehensive,
  title={A comprehensive review on software comprehension models},
  author={Fekete, Anett and Porkol{\'a}b, Zolt{\'a}n},
  booktitle={Annales Mathematicae et Informaticae},
  volume={51},
  pages={103--111},
  year={2020},
  organization={Liceum University Press}
}

@article{singh2024empirical,
  title={An empirical approach to understand the role of emotions in code comprehension},
  author={Singh, Divjot and Mishra, Ashutosh and Aggarwal, Ashutosh},
  journal={Journal of Computer Languages},
  volume={79},
  pages={101269},
  year={2024},
  publisher={Elsevier}
}

@inproceedings{adeli2020supporting,
  title={Supporting code comprehension via annotations: Right information at the right time and place},
  author={Adeli, Marjan and Nelson, Nicholas and Chattopadhyay, Souti and Coffey, Hayden and Henley, Austin and Sarma, Anita},
  booktitle={2020 IEEE symposium on visual languages and human-centric computing (VL/HCC)},
  pages={1--10},
  year={2020},
  organization={IEEE}
}

@inproceedings{middleton2024barriers,
  title={Barriers for students during code change comprehension},
  author={Middleton, Justin and Ore, John-Paul and Stolee, Kathryn T},
  booktitle={Proceedings of the IEEE/ACM 46th International Conference on Software Engineering},
  pages={1--13},
  year={2024}
}

@misc{micro_f1,
  author={Anonymous},
  title = {Micro F1 Score} ,
  note = {https://scikit-learn.org/1.5/modules/generated/sklearn.metrics.f1\_score.html}
}

@misc{rouge,
  author={Anonymous},
  title = {ROUGE Score} ,
  note = {https://pypi.org/project/rouge-score/}
}

@article{ren2020codebleu,
  title={Codebleu: a method for automatic evaluation of code synthesis},
  author={Ren, Shuo and Guo, Daya and Lu, Shuai and Zhou, Long and Liu, Shujie and Tang, Duyu and Sundaresan, Neel and Zhou, Ming and Blanco, Ambrosio and Ma, Shuai},
  journal={arXiv preprint arXiv:2009.10297},
  year={2020}
}

\end{document}